\documentclass[%
 reprint,
 amsmath,amssymb,
 jcp,
 floatfix,
]{revtex4-1}
\usepackage[dvips]{graphicx}
\usepackage{dcolumn}
\usepackage{bm}
\usepackage{verbatim}
\usepackage{color}
\usepackage{subfigure}

\renewcommand{\vec}[1]{{\mathbf{#1}}}

\begin{document}

\title{Adaptive boundaries in multiscale simulations}

\author{Jason A. Wagoner}
 \email{jason.wagoner@stonybrook.edu}
\affiliation{%
Laufer Center for Physical and Quantitative Biology, Stony Brook University, Stony Brook, New York 11794, USA
}%
\author{Vijay S. Pande}%
\affiliation{%
Department of Chemistry, Department of Structural Biology, and Department of Computer Science, Stanford University, 
Stanford, California 94305, USA
}%

\date{\today}

\begin{abstract}
Combined-resolution simulations are an effective way to study molecular properties across a range of length- and time-scales. These simulations can benefit from adaptive boundaries that allow the high-resolution region to adapt (change size and/or shape) as the simulation progresses. The number of degrees of freedom required to accurately represent even a simple molecular process can vary by several orders of magnitude throughout the course of a simulation, and adaptive boundaries react to these changes to include an appropriate but not excessive amount of detail. Here, we derive the Hamiltonian and distribution function for such a molecular simulation. We also design an algorithm that can efficiently sample the boundary as a new coordinate of the system. We apply this framework to a mixed explicit/continuum representation of a peptide in solvent. We use this example to discuss the conditions necessary for a successful implementation of adaptive boundaries that is both efficient and accurate in reproducing molecular properties.

\end{abstract}

\maketitle

Multiscale models are 
an effective way to simulate molecular systems.  The motivation is clear:
a high-resolution model can capture physical detail while a
low-resolution model offers computational efficiency and is sometimes better suited or more easily parameterized for large-scale phenomena~\cite{Noid:2013jk}.
There are two strategies for multiscale modeling.  In the first, multiple independent simulations are used for different levels of resolution.  Combined intelligently, the sum is worth more than the parts:  simulations at one level motivate and parameterize simulations at another~\cite{Ayton:2007fz, Ayton:2010fm, Lyman:2008fi}. 
The focus of this work is the second approach, which combines multiple levels of resolution into a single simulation.  These simulations use a fine-grained model for a region of interest and a computationally efficient coarse-grained model elsewhere.  Examples include mixed quantum and molecular mechanical~\cite{Gao:1998wk,Warshel:2003ip}, mixed all-atom and coarse grained (CG)~\cite{Villa:2004jg, Ayton:2007ck,Wassenaar:2013gv,Praprotnik:2007gya,Zavadlav:2014bj,Fogarty:2015bg,Zavadlav:2015kh}, and hybrid explicit-continuum solvent models~\cite{Beglov:1994ip,Wagoner:2011fp,Wagoner:2013kj,Deng:2008cd,Brunger:1984cv,Im:2001ha,Lee:2004hg,Petsev:2015fm}.

 Accurately modeling the boundary between high- and low-resolution regions is the crux of a combined-resolution simulation. 
 Even for a homogeneous system like bulk solvent, equilibrium properties emerge from a delicate balance of interactions with the surrounding medium. An improper handling of the boundary will break the natural symmetry, and the resulting structural artifacts can extend well beyond the boundary into other regions of the simulation~\cite{Wagoner:2011fp}.
 
  We have developed a hybrid explicit-continuum solvent model that includes a boundary region over which molecules gradually, rather than abruptly, change resolution. This boundary method avoids the structural artifacts common to hybrid solvent models and accurately reproduces thermodynamic properties throughout the entire explicit domain~\cite{Wagoner:2011fp,Wagoner:2013kj}.
 This boundary method is similar to that introduced in the Adaptive Resolution (AdRes) approach~\cite{Praprotnik:2005eq,Praprotnik:2008er}, a method used to couple high resolution particles to a more coarse-grained~\cite{Wassenaar:2013gv,Praprotnik:2007gya} or continuum~\cite{DelgadoBuscalioni:2008iw,DelgadoBuscalioni:2009gs,Walther:2012gv} representation and that can also incorporate regions with some quantum mechanical effects~\cite{Poma:2010hf,Agarwal:2015fx,Poma:2011ct}. AdRes has been successfully applied to a range of molecular systems~\cite{Wassenaar:2013gv,Praprotnik:2007gya,Zavadlav:2014bj,Fogarty:2015bg,Zavadlav:2015kh,Krekeler:2017ga}.

\begin{figure}
 \centering
  \includegraphics[width=8.6cm]{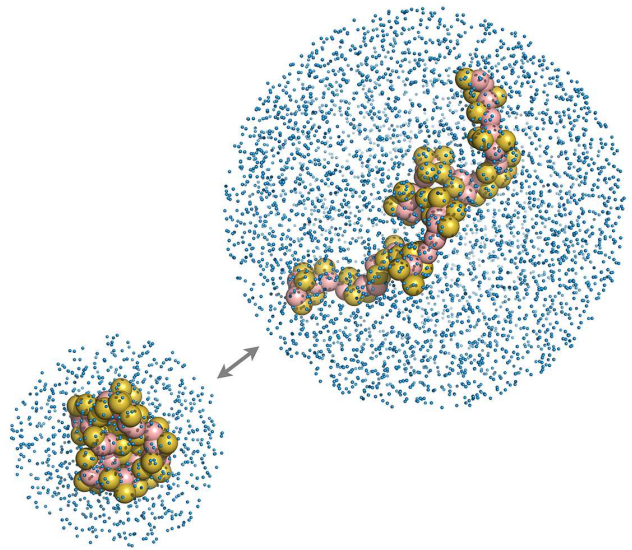}
\caption{\label{F:SphereUpdate} {A mixed explicit/continuum solvent model with an adaptive boundary that shrinks and expands in response to the conformational fluctuations of the peptide. A video of this simulation has been posted online~\cite{50merMovie}.}}
\end{figure}

  Previously~\cite{Wagoner:2013kj},
we introduced adaptive boundaries into combined-resolution simulations. 
 This allows the high-resolution region to adapt and include an appropriate but not excessive amount of detail as the simulation progresses.  
 Consider protein folding as a simple example. 
 A simulation with a fixed boundary must be large enough to solvate the largest possible protein conformation. 
 As we show below, an infrequently-visited extended conformation may require over an order of magnitude more solvent molecules than the predominant collapsed state. In contrast, an adaptive boundary can shrink and expand as the simulation progresses, as shown in Figure \ref{F:SphereUpdate}.  Adaptive boundaries would similarly benefit simulations of biomolecular assembly, aggregation, crystallization, and other examples outside of biology.

Simulations with flexible boundaries have been previously implemented with a restraining potential centered on a solute molecule that prevents atomistic solvent from drifting away~\cite{Li:2006hx,Szklarczyk:2015in}. Kreis \textit{et. al.} have implemented an adjustable boundary with AdRes, which does not use a restraining potential and defines the high resolution region by a set of overlapping spheres that can change relative position~\cite{Kreis:2016hl}.
This method successfully reproduces thermodynamic properties. 

 Here, we define the Hamiltonian and derive the distribution function of a combined-resolution simulation that has adaptive boundaries.  
 This approach complements the work of Kreis \textit{et. al.}, which does not have a conserved Hamiltonian. Their method could presumably be incorporated with the Hamiltonian-based version of AdRes~\cite{Potestio:2013hwa,Kreis:2014bx,Espanol:2015cl,Site:2017kf}, in which case the appropriate distribution function should match that derived here. 

  This formalism allows us to connect the equilibrium properties of the high resolution region to those expected from a simulation performed in full detail.  We build on previous work~\cite{Wagoner:2013kj}, where we derived adaptive boundaries specifically for a mixed explicit-continuum model with a spherical domain. Our work here extends this theory for general mixed-resolution models and for arbitrarily shaped boundaries.  
We also design a new algorithm that efficiently samples the boundary as a coordinate of the system.  We test this model on a peptide in a mixed explicit/continuum solvent model, as shown in Figure \ref{F:SphereUpdate}, and show that an adaptive boundary severely reduces the number of degrees of freedom in the simulation. 

 We first outline the theory of combined-resolution models with fixed boundaries following the formalism of Roux and coworkers~\cite{Beglov:1994ip,Im:2001ha,Deng:2008cd}. 
Consider a system containing molecule $A$ of interest (e.g., a protein), which is always modeled in high resolution, and $N$ identical solvent molecules. 
We define fine-grained $U$ and coarse-grained $V$ potential functions.   
This is a minimal example; the theory is easily extended to multi-component systems or to a molecule $A$ that can change resolution. 

We first consider the full system in fine-grained (FG) detail.  The configurational probability distribution in the canonical ensemble is
\begin{equation}
P^{\text{FG}} \left( \vec{X} \right) = \frac{1}{Z^{\text{FG}}}\mbox{exp} \left[ -\beta H^{\text{FG}} ( \vec{X})\right] ,
\label{e:FGCanonicalProb}
\end{equation}
\begin{equation}
Z^{\text{FG}} = \frac{1}{N!}\int_{\Omega} d \vec{X}_A \int_{\Omega} d \vec{X}_N \mbox{exp} \left[ -\beta H^{\text{FG}} ( \vec{X} )\right],
\label{e:FGZ}
\end{equation}
for configuration $\vec{X}$, where the subscripts $A$ and $N$ correspond to molecule $A$ and the $N$ solvent molecules.  
Henceforth, we will drop the coordinate arguments of the Hamiltonian. 
The Hamiltonian can be written as a sum of intra- and inter-molecular terms;
$H^{\text{FG}}= \sum_i  U_{i}  +\sum_{i,j \neq i} U_{ij} /2$ for 
$ i,j \in \left\{ A,1,\ldots,N \right\}$. 

Now consider the same system modeled in combined-resolution with a fixed boundary. 
We partition the domain into regions of high- and low-resolution,
$\Omega = \Omega_{\text{i}} \left(\Gamma \right)  \cup \Omega_{\text{o}} \left(\Gamma \right) $,
delineated by the boundary $\Gamma$.  The representation of molecule $i$ is defined by the scaling function $\lambda_i = \lambda \left( \vec{x}_i, \Gamma  \right)$.   
Molecules in high resolution correspond to $\lambda_i = 1$ for $\vec{x}_i \in \Omega_{\text{i}}$. Molecules in low resolution correspond to 
$\lambda_i = 0$.  This switch may occur abruptly, so that $\lambda_i = 0$ for $\vec{x}_i \in \Omega_{\text{o}}$, or $\lambda_i$ may smoothly interpolate to zero over some transition region~\cite{Wagoner:2011fp,Wagoner:2013kj,Praprotnik:2007gya,Wassenaar:2013gv}.  
The multiscale Hamiltonian is
\begin{eqnarray}
H^{\text{MM}} \left( \Gamma \right) &=&
\Delta W + 
\sum_{i  }
\left[ 
\lambda_i U_{i}+\left( 1 -  \lambda_{i} \right)  V_{i} 
\right. \nonumber \\ & & \left.
+
\frac{1}{2}\sum_{j \neq i}
\left(
\lambda_{i}\lambda_j U_{ij} + 
\left( 1 - \lambda_i \lambda_j \right) V_{ij} 
\right) \right],
\label{e:Hmm}
\end{eqnarray}
where $\Delta W = \Delta W \left( \vec{X}, \Gamma \right) $ is a many-body potential of mean force (PMF). 
It is implied that $V$ is a compound function that contains a mapping $M$ to low resolution, $M: \vec{X} \rightarrow \vec{Y}$, where $\vec{Y}$ is a coarse-grained representation of $\vec{X}$. 
  The configurational probability distribution of this combined-resolution system is
\begin{equation}
P^{\text{MM-fixed}} \left( \vec{X} ,\Gamma \right) = \mbox{exp} \left[ -\beta H^{\text{MM} }
\left( \Gamma \right)  \right] / Z^{\text{MM}}. 
\label{e:CanonicalMM}
\end{equation}
Where the superscript signals that the boundary is fixed, though  $H^{\text{MM}}$ depends on its location $\Gamma$.

To reproduce the  thermodynamic properties of the interior region, 
we define the marginal probability distribution of the high-resolution system:
\begin{eqnarray}
P^{\text{FG}} \left( \vec{X}_n, \Gamma \right) &=&
\frac{\mbox{exp} \left[ -\beta  H^{\text{FG}}_{\text{ii}}  \right]}
{ n! (N-n)! Z^{\text{FG}} }
\label{e:FGreduced} \\ & & \times
 \int_{\Omega_{\text{o}} \left( \Gamma \right)} 
 d \vec{X}_{N-n} \mbox{exp} \left[ -\beta  \left( H^{\text{FG}}_{\text{io}} 
 + 
 H^{\text{FG}}_{\text{oo}} 
 \right) \right],
\nonumber
\end{eqnarray}
and aim to reproduce this function with the analogous marginal distribution of the combined-resolution system.   
We have separated the potential into inner-inner, $H^{\text{FG}}_{\text{ii}}$, inner-outer, $H^{\text{FG}}_{\text{io}}$, and outer-outer,  $H^{\text{FG}}_{\text{oo}}$, components~\cite{Wagoner:2011fp} and the subscripts $n$ and $N-n$ denote molecules located within the interior ($\Omega_{\text{i}}$) and exterior ($\Omega_{\text{o}}$) regions.   
If we do the same for equation \ref{e:CanonicalMM}, we can ensure $P^{\text{MM-fixed}}\left( \vec{X}_n,\Gamma \right) = P^{\text{FG}}\left( \vec{X}_n, \Gamma \right) $ by imposing 
\begin{eqnarray} && 
 \int_{\Omega_{\text{o}} \left( \Gamma \right)} 
 d \vec{X}_{N-n} \mbox{exp} 
 \left[ -\beta  \left( H^{\text{FG}}_{\text{io}} +
 H^{\text{FG}}_{\text{oo}}   \right) \right]
 = 
 \label{e:PMFequiv} \\ & & \qquad \qquad
  \int_{\Omega_{\text{o}} \left( \Gamma \right)} 
 d \vec{X}_{N-n} \mbox{exp} \left[ -\beta  \left( H^{\text{MM}}_{\text{io}} 
 + 
 H^{\text{MM}}_{\text{oo}}  + \Delta W 
 \right) \right].
\nonumber 
 \end{eqnarray}
 The goal in parameterizing a coarse-grained potential $V$ and/or PMF $\Delta W$ is to satisfy equation~\ref{e:PMFequiv}~\cite{Beglov:1994ip, Wagoner:2011fp}.

We can now explicitly include adaptive boundaries in the distribution function. We define a normalized joint probability $P \left( \vec{X}_A,\Gamma \right)$ that couples the boundary location to the configuration of molecule \emph{A} and modify equation \ref{e:CanonicalMM} to include the boundary as an explicit coordinate: 
\begin{equation}
P^{\text{MM}}\left( \vec{X}, \Gamma \right)= 
\frac{1}{Z^{\text{MM}}}
P \left( \vec{X}_A,\Gamma \right) 
\mbox{exp} \left[ -\beta  H^{\text{MM}} \right].
\label{e:Pmm}
\end{equation}
Equation~\ref{e:Pmm} is the distribution function sampled from a multiscale simulation with adaptive boundaries. 
Consider some molecule or molecules always given in high-resolution (here, molecule $A$). 
This distribution will exactly reproduce the thermodynamic properties for these molecules
if (1) equation~\ref{e:PMFequiv} is satisfied, and (2) $P \left( \vec{X}_A,\Gamma \right)$  is normalized, as we can see by integrating over the boundary $\Gamma$ and all other degrees of freedom when these conditions are satisfied: 
\begin{eqnarray}
P^{\text{MM}}\left( \vec{X}_A \right) & = & 
\frac{1}{Z^{\text{MM}}}
 \int d \Gamma P \left( \vec{X}_A,\Gamma \right) 
  \int d \vec{X}_N \mbox{exp} \left[ -\beta  H^{\text{MM}} \right]
 \nonumber \\ & = &
 \frac{1}{Z^{\text{FG}}}  \int 
 d \vec{X}_N \mbox{exp} \left[ -\beta H^{\text{FG}} ( \vec{X})\right]
\nonumber \\ & = & P^{\text{FG}} \left( \vec{X}_A \right). 
 \label{e:PmmA}
\end{eqnarray}
Equations~\ref{e:Pmm}-\ref{e:PmmA} give our first main result.

For a static domain, the $\Gamma$-dependence of $\Delta W$ may be safely ignored.  To include adaptive boundaries, it is essential that $\Delta W$ accurately capture this dependence.  For a hybrid explicit-continuum model, for example, $\Delta W$ must have a cavitation term that is very accurate with respect to the shape and size of the explicit domain. Otherwise, the domain of the simulation will tend toward large or small sizes and, through equation~\ref{e:Pmm}, artificially bias the configuration of molecule $A$~\cite{Wagoner:2013kj}.

The low-resolution region can be represented through a number of CG models~\cite{Shell:2008cj,Chaimovich:2010ic,Shell:2012kj,Mullinax:2009hs,Mullinax:2010cz,Izvekov:2005ig,Noid:2008dc,Das:2009il,Dama:2013bm,Davtyan:2014gq,Dama:2013bm,Davtyan:2014gq} or with a pure continuum~\cite{Roux:1999gf,Beglov:1994ip,Brunger:1984cv,Im:2001ha,Lee:2004hg}. A combination of these two limiting cases works well, and we have developed a model that includes a `flawed region' of solvent molecules that  gradually transition from explicit detail to continuum. This flawed region reproduces local interactions and relaxes the complexity of $\Delta W$~\cite{Wagoner:2013kj}.  
 The scaling function is
   \begin{equation}
\lambda_i = \begin{cases}
1 & r_i \leq R-w \\
1 +\frac{2\left(r_i-R+w\right)^3}{w ^3} 
-\frac{3\left(r_i-R+w\right)^2}{w^2} & 
R-w < r_i < R  \\
0 & r_i \geq R  
\end{cases},
\label{e:Spline}
\end{equation}
   where $r_i$ is the distance of molecule $i$ from the center of the domain, $R$ is the location of the boundary, and $w$ defines the width of the transition region.

In our model, the PMF $\Delta W$ includes a position-dependent chemical potential that accounts for transforming molecules from an FG to continuum representation and ensures constant solvent density across the transition region~\cite{Wagoner:2011fp,Wagoner:2013kj}.  The multiscale Hamiltonian, equation~\ref{e:Hmm}, leads to position-dependent forces that cannot be written as a sum of pairwise antisymmetric terms between molecules. These forces arise from interactions between a molecule and the degrees of freedom that have been `removed' from the system and are also present in other hybrid models that rely on a PMF construction~\cite{Beglov:1994ip,Im:2001ha,Deng:2008cd,Lee:2004hg}.  
This approach gives a sound thermodynamic formalism~\cite{Beglov:1994ip,Wagoner:2011fp,Wagoner:2013kj,Everaers:2016dw} and matches the Hamiltonian version of AdRes~\cite{Potestio:2013hwa,Kreis:2014bx,Espanol:2015cl,Site:2017kf} though Newton's third law is not conserved between molecules.

Alternatively, the force-based version of AdRes is constructed to have forces that are a sum of pairwise antisymmetric terms~\cite{Praprotnik:2005eq,Praprotnik:2008er,DelgadoBuscalioni:2008iw,Kreis:2014bx}, important to describe, for example, hydrodynamics at the boundary~\cite{DelleSite:2007fu,DelgadoBuscalioni:2008iw,Kreis:2014bx,Site:2017kf}. This method does not have a conserved Hamiltonian but does conserve Newton's third law between molecules.  We do not consider hydrodynamics here since it would not affect thermodynamics and, as we discuss below, does not seem to be necessary for the kinetics of biomolecules in our model. Hydrodynamics can be important for other systems and 
have been considered in more detail 
in AdRes~\cite{DelgadoBuscalioni:2008iw,Kreis:2014bx,Site:2017kf} and other hybrid models~\cite{Flekkoy:2000ti,Giupponi:2007br,Korotkin:2015go,Petsev:2015fm}.

We now apply this work to simulations of a peptide in a sphere of explicit solvent.
We use a boundary $\Gamma = R$ that defines the inner region $\Omega_{\text{i}}$ as a sphere of radius $R$ that is coupled to the radius of gyration $R_g$ of the peptide using a Gaussian distribution:  
\begin{equation}
P \left( \vec{X}_A, R \right) = \sqrt{\frac{\beta k}{\pi}} \mbox{exp}
\left[  -\beta k \left( R - a_R - b_R R_g \left( \vec{X}_A \right)  \right)^2   \right], 
\label{e:RgRsCoupling}
\end{equation}
where $a_R$ and $b_R$ determine the amount of space between the peptide and boundary,
 and $k$ is the coupling strength.  

\begin{figure}
 \centering
  \includegraphics[width=8.6cm]{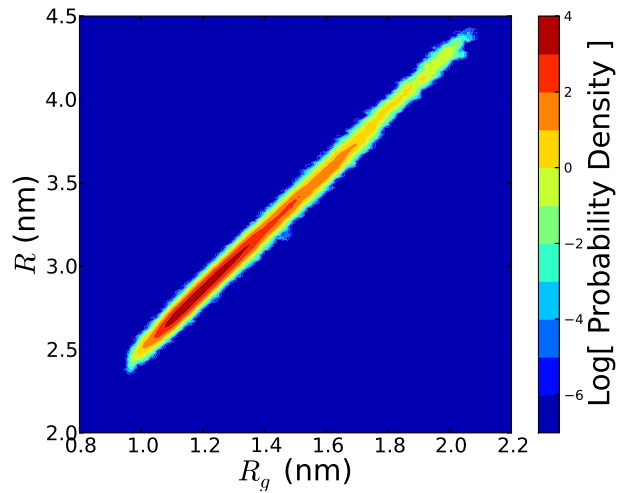}
\caption{\label{F:RGRS} {Log joint probability distribution of the explicit sphere radius $R$ and the radius of gyration $R_g$ for a 50-residue repeat of polyglutamine.  The adaptive boundary of the sphere shrinks and expands in response to the conformational fluctuations of the peptide. 
 }}
\end{figure}

Our algorithm has two steps: (1) the configuration $ \vec{X} $ is updated with a combination of molecular dynamics and grand canonical Monte Carlo (MC) moves; (2) the boundary is updated with an MC move.
Previously, we proposed updating the boundary to discrete positions using MC moves biased by nonequilibrium paths that required hundreds of integration steps~\cite{Wagoner:2013kj} per boundary update. 
The following method is more efficient and more easily applied. 
We instead define the boundary as a continuous variable and modify the Hamiltonian so that boundary update can be completed without any integration steps. 
We define two bounding radii, $\underline{R} = R_h \left \lfloor \frac{R}{R_h} \right \rfloor$
and $\overline{R} = R_h \left \lceil \frac{R}{R_h} \right \rceil$, 
 for a given discretization $R_h$
 with the floor $\left \lfloor \right \rfloor$  and ceiling $\left \lceil \right \rceil$ functions. 
 The new Hamiltonian is a linear combination of the energies corresponding to the two bounding radii:
\begin{equation}
H_{\alpha} \left( \vec{X} ,  R \right) = \left( 1 - \alpha\right) 
H^{\text{MM}} \left( \vec{X}, \underline{R} \right) 
 +
\alpha 
H^{\text{MM}} \left( \vec{X} , \overline{R} \right),
\label{e:Hmm.Scaled}
\end{equation}
where  $\alpha \left( R \right) = \left( R - \underline{R} \right) / R_h$. 
 The two energies on the right hand side of equation \ref{e:Hmm.Scaled} can be calculated efficiently and simultaneously as they differ only in their interactions for molecules in the transition region. 
 
   For the MC move, we hold all molecules fixed and select a candidate radius $R'$ from a small uniform window (0.01 nm) centered on the current value $R$.  We accept or reject the move with a Metropolis criterion.  If the candidate radius $R'$ does not cross one of the bounding radii, then $H^{\text{MM}} \left( \vec{X}, \underline{R} \right)$ and $H^{\text{MM}} \left( \vec{X} , \overline{R} \right)$ do not change and the new energy is calculated from equation \ref{e:Hmm.Scaled} by updating the value of $\alpha$. 
    This move is of negligible computational expense. Alternatively, the candidate boundary $R'$ will cross one of the bounding radii. If $R' > \overline{R}$, we must insert a new shell of molecules.  At the moment that $R' = \overline{R}$, these new molecules are non-interacting and the candidate solvent shell is inserted using the distribution of ideal molecules.
Because the candidate $R'$ will not fall precisely on $\overline{R}$, this move will have a small energetic contribution calculated from equation \ref{e:Hmm.Scaled}. 
Similarly, for a move that crosses the lower bounding radius, $R' < \underline{R}$, we automatically delete the appropriate shell of water molecules.  The selection probabilities of the insertion and deletion moves cancel exactly with their contributions to the overall configurational probability distribution~\cite{Wagoner:2013kj}. This algorithm is our second main result.

This scheme leverages the weak interactions at the boundary to construct a new high-resolution configuration from the previous low-resolution configuration. Because there is a gradual change in resolution, these new coordinates are close to equilibrium and we have high acceptance rates. This would be more difficult with an abrupt change in resolution at the boundary.

We now test the adaptive boundary algorithm on simulations of solvated peptides. 
We use the MARTINI forcefield~\cite{Marrink:2007bw,Monticelli:2008ia,Wagoner:2013kj}, a coarse representation that does not include solvent charged interactions.
 Simulations are preformed with a Langevin integrator~\cite{Bussi:2007jx} and a generalized hybrid MC correction for the finite time-step~\cite{Wagoner:2012fe}. We set the width of the transition region defined in equation~\ref{e:Spline} to $w=0.5$ nm, roughly the size of a single solvation layer in MARTINI. This value has been found to accurately reproduce thermodynamic properties~\cite{Wagoner:2011fp}. After testing multiple parameter sets for the boundary location defined in equation~\ref{e:RgRsCoupling}, we set $a_R= 0.7$ nm, $b_R =$ 1.8, and $k=2000$ kJ/mol/nm$^2$. These values give a tight shell of 1-3 solvent layers around the peptide throughout the course of the simulation, though different parameters may be suitable for other biomolecules. 
The bulk PMF $\Delta W$ includes a term, $w_{\text{cav}} \left(R \right) = a R^3 + bR^2 + cR$, where $a$ = 207.20 kJ/mol/nm$^3$, $b$ = -9.00 kJ/mol/nm$^2$, and $c$ = $-9.17$ kJ/mol/nm.  This term models the change in $\Delta W$ as a function of the boundary position and is essential to ensure that the explicit domain is not artificially biased to large or small sizes. Parameterization of this and other contributions to $\Delta W$ are described in Ref.~\cite{Wagoner:2013kj}.  

The adaptive boundary method is illustrated in Figure~\ref{F:SphereUpdate} for a simulation of a 50-residue repeat of polyglutamine. 
The joint distribution of $R_g$ and the sphere radius $R$ is shown in Figure \ref{F:RGRS}.  The number of explicit particles, proportional to volume, spans from 440 to 2800, a sixfold difference over the course of the simulation. Were this simulation performed in atomistic detail, rather than with MARTINI (where one particle corresponds to 4 water molecules), the corresponding range would be 72-fold.
While [688, 996] solvent molecules are required for the interquartile range, the infrequently visited states ($R_g > 1.72 $ nm, ~2\% of the simulation) require more than 1880 molecules.   This shows the computational advantage of the method: the size of the high-resolution region can remain small unless more detail is needed. 

\begin{figure}
 \centering
  \includegraphics[width=8.6cm]{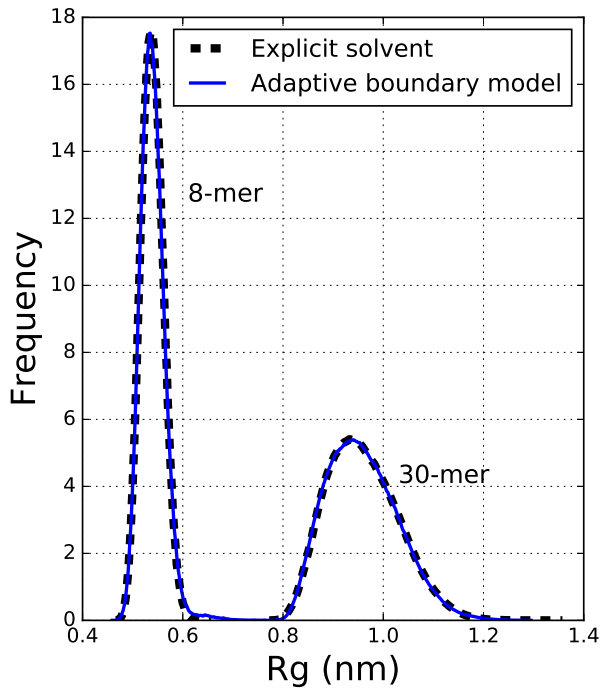}
\caption{\label{F:RgComparison} {The distribution of $R_g$ for an 8- and 30-residue repeat of polyglutamine calculated from full explicit simulations and from the adaptive boundary multiscale model.  Results of the adaptive boundary method are exact within statistical error.}}
\end{figure}

This model reproduces thermodynamic properties exactly, within statistical error, as shown in Figure \ref{F:RgComparison}.  This is the distribution of $R_g$ compared to a pure explicit simulation for a 30-residue and an 8-residue repeat of polyglutamine.  We have obtained this degree of accuracy for multidimensional thermodynamic properties, such as the principal moments of inertia (data not shown).  

Our derivation of $H^{\text{MM}}$ is committed to reproducing thermodynamic properties but gives no consideration to kinetics.  The integrated autocorrelation function for $R_g$ of the polyglutamine octamer shows an increase from 38.33 ps in full explicit detail to 44.34 ps in the adaptive boundary model.  The 16\% increase is unsurprising: kinetics may be affected by both the continuum representation and lack of hydrodynamics for solvent in the outer domain and by the introduction of MC boundary updates.  We find that the former effect is negligible here by testing the model over larger explicit regions. 
 It may be possible to develop boundary updates with memory in order to reduce this small error in kinetics. 

In summary, we have derived the distribution function and have given an example implementation of adaptive boundaries for combined-resolution simulations.  
In our example, the boundary reacts to the peptide's radius of gyration to
include an appropriate but not excessive amount of detail as the simulation progresses. 
  Adaptive boundaries would similarly benefit simulations of crystallization, aggregation, biomolecular assembly, and others. 

Our approach should transfer to other models that use smooth coupling at the boundary. 
An adaptive boundary simulation requires both a very accurate model of the bulk PMF as a function of the system's boundary and a way to generate equilibrium high-resolution configurations mapped from low-resolution configurations. Both of these requirements can be met using a smoothly-coupled boundary that has a gradual change in resolution. 

The example we have used here is simple and changes the size but not the shape of a spherical domain.  Kreis \textit{et. al.} have implemented a domain of several overlapping spheres that cannot change size, but do adjust shape by changing their relative orientations~\cite{Kreis:2016hl}.
This approach could be included here to give a more flexible domain that can change both size and shape. 
  With fixed boundaries, this model can include electrostatic interactions using Generalized Born-like terms with volume overlaps~\cite{Grycuk:2003fw} or  a reaction-field term solved for a smoothly-varying dielectric~\cite{Qin:2009ej,Xue:2010ih}. For adaptive boundaries, the remaining challenge is to parameterize the boundary-dependence of these electrostatic contributions to the PMF. 

\vspace{6mm}
\section*{Acknowledgements}

We are grateful to Sean Seyler for a discussion on the role of hydrodynamic interactions in combined-resolution models. We thank Emiliano Brini  and James Dama for helpful comments on this manuscript.  
This work was supported by NIH award GM62868.  


\begin{thebibliography}{10}

\bibitem{Noid:2013jk}
W.~G. Noid,
\newblock J. Chem. Phys. {\bf 139}, 090901 (2013).

\bibitem{Ayton:2007fz}
G.~S. Ayton, W.~G. Noid, and G.~A. Voth,
\newblock Curr. Opin. Struct. Biol. {\bf 17}, 192 (2007).

\bibitem{Ayton:2010fm}
G.~S. Ayton and G.~A. Voth,
\newblock Biophys. J. {\bf 99}, 2757 (2010).

\bibitem{Lyman:2008fi}
E.~Lyman, J.~Pfaendtner, and G.~A. Voth,
\newblock Biophys. J. {\bf 95}, 4183 (2008).

\bibitem{Gao:1998wk}
J.~Gao, P.~Amara, C.~Alhambra, and M.~J. Field,
\newblock J. Phys. Chem. A {\bf 102}, 4714 (1998).

\bibitem{Warshel:2003ip}
A.~Warshel,
\newblock Annu. Rev. Biophys. {\bf 32}, 425 (2003).

\bibitem{Villa:2004jg}
E.~Villa, A.~Balaeff, L.~Mahadevan, and K.~Schulten,
\newblock Multiscale Model Simul. {\bf 2}, 527 (2004).

\bibitem{Ayton:2007ck}
G.~S. Ayton and G.~A. Voth,
\newblock J. Struct. Biol. {\bf 157}, 570 (2007).

\bibitem{Wassenaar:2013gv}
T.~A. Wassenaar, H.~I. Ing{\'o}lfsson, M.~Prie{\ss}, S.~J. Marrink, and L.~V.
  Sch{\"a}fer,
\newblock J. Phys. Chem. B {\bf 117}, 3516 (2013).

\bibitem{Praprotnik:2007gya}
M.~Praprotnik, S.~Matysiak, L.~D. Site, K.~Kremer, and C.~Clementi,
\newblock J. Phys. Condens. Matter {\bf 19}, 292201 (2007).

\bibitem{Zavadlav:2014bj}
J.~Zavadlav, M.~N. Melo, S.~J. Marrink, and M.~Praprotnik,
\newblock J. Chem. Phys. {\bf 140}, 054114 (2014).

\bibitem{Fogarty:2015bg}
A.~C. Fogarty, R.~Potestio, and K.~Kremer,
\newblock J. Chem. Phys. {\bf 142}, 195101 (2015).

\bibitem{Zavadlav:2015kh}
J.~Zavadlav, R.~Podgornik, and M.~Praprotnik,
\newblock J. Chem. Theory Comput. {\bf 11}, 5035 (2015).

\bibitem{Beglov:1994ip}
D.~Beglov and B.~Roux,
\newblock J. Chem. Phys. {\bf 100}, 9050 (1994).

\bibitem{Wagoner:2011fp}
J.~A. Wagoner and V.~S. Pande,
\newblock J. Chem. Phys. {\bf 134}, 214103 (2011).

\bibitem{Wagoner:2013kj}
J.~A. Wagoner and V.~S. Pande,
\newblock J. Chem. Phys. {\bf 139}, 234114 (2013).

\bibitem{Deng:2008cd}
Y.~Deng and B.~Roux,
\newblock J. Chem. Phys. {\bf 128}, 115103 (2008).

\bibitem{Brunger:1984cv}
A.~Br{\"u}nger, C.~L. Brooks, and M.~Karplus,
\newblock Chem. Phys. Lett. {\bf 105}, 495 (1984).

\bibitem{Im:2001ha}
W.~Im, S.~Bern{\`e}che, and B.~Roux,
\newblock J. Chem. Phys. {\bf 114}, 2924 (2001).

\bibitem{Lee:2004hg}
M.~S. Lee, F.~R. Salsbury, and M.~A. Olson,
\newblock J. Comput. Chem. {\bf 25}, 1967 (2004).

\bibitem{Petsev:2015fm}
N.~D. Petsev, L.~G. Leal, and M.~S. Shell,
\newblock J. Chem. Phys. {\bf 142}, 044101 (2015).

\bibitem{Praprotnik:2005eq}
M.~Praprotnik, L.~Delle~Site, and K.~Kremer,
\newblock J. Chem. Phys. {\bf 123}, 224106 (2005).

\bibitem{Praprotnik:2008er}
M.~Praprotnik, L.~D. Site, and K.~Kremer,
\newblock Annu. Rev. Phys. Chem. {\bf 59}, 545 (2008).

\bibitem{DelgadoBuscalioni:2008iw}
R.~Delgado-Buscalioni, K.~Kremer, and M.~Praprotnik,
\newblock J. Chem. Phys. {\bf 128}, 114110 (2008).

\bibitem{DelgadoBuscalioni:2009gs}
R.~Delgado-Buscalioni, K.~Kremer, and M.~Praprotnik,
\newblock J. Chem. Phys. {\bf 131}, 244107 (2009).

\bibitem{Walther:2012gv}
J.~H. Walther, M.~Praprotnik, E.~M. Kotsalis, and P.~Koumoutsakos,
\newblock J. Comput. Phys. .

\bibitem{Poma:2010hf}
A.~B. Poma and L.~Delle~Site,
\newblock Phys. Rev. Lett. {\bf 104}, 250201 (2010).

\bibitem{Agarwal:2015fx}
A.~Agarwal and L.~Delle~Site,
\newblock J. Chem. Phys. {\bf 143}, 094102 (2015).

\bibitem{Poma:2011ct}
A.~B. Poma and L.~D. Site,
\newblock Phys. Chem. Chem. Phys. {\bf 13}, 10510 (2011).

\bibitem{Krekeler:2017ga}
C.~Krekeler and L.~D. Site,
\newblock Phys. Chem. Chem. Phys. {\bf 19}, 4701 (2017).

\bibitem{Li:2006hx}
Y.~Li, G.~Krilov, and B.~J. Berne,
\newblock J. Phys. Chem. B {\bf 110}, 13256 (2006).

\bibitem{Szklarczyk:2015in}
O.~M. Szklarczyk, N.~S. Bieler, P.~H. H{\"u}nenberger, and W.~F. van Gunsteren,
\newblock J. Chem. Theory Comput. {\bf 11}, 5447 (2015).

\bibitem{Kreis:2016hl}
K.~Kreis, R.~Potestio, K.~Kremer, and A.~C. Fogarty,
\newblock J. Chem. Theory Comput. {\bf 12}, 4067 (2016).

\bibitem{Potestio:2013hwa}
R.~Potestio {\em et~al.},
\newblock Phys. Rev. Lett. {\bf 110}, 108301 (2013).

\bibitem{Kreis:2014bx}
K.~Kreis, D.~Donadio, K.~Kremer, and R.~Potestio,
\newblock Europhys. Lett. {\bf 108}, 30007 (2014).

\bibitem{Espanol:2015cl}
P.~Espa{\~n}ol {\em et~al.},
\newblock J. Chem. Phys. {\bf 142}, 064115 (2015).

\bibitem{Site:2017kf}
L.~D. Site and M.~Praprotnik,
\newblock Phys. Rep. {\bf 693}, 1 (2017).

\bibitem{50merMovie}
\url{http://y2u.be/wgEnF9BCNW0}

\bibitem{Shell:2008cj}
M.~S. Shell,
\newblock J. Chem. Phys. {\bf 129}, 144108 (2008).

\bibitem{Chaimovich:2010ic}
A.~Chaimovich and M.~S. Shell,
\newblock Phys. Rev. E {\bf 81}, 060104 (2010).

\bibitem{Shell:2012kj}
M.~S. Shell,
\newblock J. Chem. Phys. {\bf 137}, 084503 (2012).

\bibitem{Mullinax:2009hs}
J.~W. Mullinax and W.~G. Noid,
\newblock Phys. Rev. Lett. {\bf 103}, 198104 (2009).

\bibitem{Mullinax:2010cz}
J.~W. Mullinax and W.~G. Noid,
\newblock J. Phys. Chem. C {\bf 114}, 5661 (2010).

\bibitem{Izvekov:2005ig}
S.~Izvekov and G.~A. Voth,
\newblock J. Phys. Chem. B {\bf 109}, 2469 (2005).

\bibitem{Noid:2008dc}
W.~G. Noid {\em et~al.},
\newblock J. Chem. Phys. {\bf 128}, 244114 (2008).

\bibitem{Das:2009il}
A.~Das and H.~C. Andersen,
\newblock J. Chem. Phys. {\bf 131}, 034102 (2009).

\bibitem{Dama:2013bm}
J.~F. Dama {\em et~al.},
\newblock J. Chem. Theory Comput. {\bf 9}, 2466 (2013).

\bibitem{Davtyan:2014gq}
A.~Davtyan, J.~F. Dama, A.~V. Sinitskiy, and G.~A. Voth,
\newblock J. Chem. Theory Comput. {\bf 10}, 5265 (2014).

\bibitem{Roux:1999gf}
B.~Roux and T.~Simonson,
\newblock Biophys. Chem. {\bf 78}, 1 (1999).

\bibitem{Everaers:2016dw}
R.~Everaers,
\newblock Eur. Phys. J. {\bf 225}, 1483 (2016).

\bibitem{DelleSite:2007fu}
L.~Delle~Site,
\newblock Phys. Rev. E {\bf 76}, 047701 (2007).

\bibitem{Flekkoy:2000ti}
E.~Flekkoy, P.~Coveney, and G.~De~Fabritiis,
\newblock Phys. Rev. E {\bf 62}, 2140 (2000).

\bibitem{Giupponi:2007br}
G.~Giupponi, G.~De~Fabritiis, and P.~V. Coveney,
\newblock J. Chem. Phys. {\bf 126}, 154903 (2007).

\bibitem{Korotkin:2015go}
I.~Korotkin {\em et~al.},
\newblock J. Chem. Phys. {\bf 143}, 014110 (2015).

\bibitem{Marrink:2007bw}
S.~J. Marrink, H.~J. Risselada, S.~Yefimov, D.~P. Tieleman, and A.~H. de~Vries,
\newblock J. Phys. Chem. B {\bf 111}, 7812 (2007).

\bibitem{Monticelli:2008ia}
L.~Monticelli {\em et~al.},
\newblock J. Chem. Theory Comput. {\bf 4}, 819 (2008).

\bibitem{Bussi:2007jx}
G.~Bussi and M.~Parrinello,
\newblock Phys. Rev. E {\bf 75}, 2289 (2007).

\bibitem{Wagoner:2012fe}
J.~A. Wagoner and V.~S. Pande,
\newblock J. Chem. Phys. {\bf 137}, 214105 (2012).

\bibitem{Grycuk:2003fw}
T.~Grycuk,
\newblock J. Chem. Phys. {\bf 119}, 4817 (2003).

\bibitem{Qin:2009ej}
P.~Qin, Z.~Xu, W.~Cai, and D.~Jacobs,
\newblock Commun. Comput. Phys. {\bf 6}, 955 (2009).

\bibitem{Xue:2010ih}
C.~Xue and S.~Deng,
\newblock Phys. Rev. E {\bf 81}, 37 (2010).

\end{thebibliography}

\end{document}